\documentclass[journal]{vgtc}                % final (journal style)
\ifpdf%                                % if we use pdflatex
  \pdfoutput=1\relax                   % create PDFs from pdfLaTeX
  \usepackage{graphicx}                % allow us to embed graphics files
  \DeclareGraphicsExtensions{.pdf,.png,.jpg,.jpeg} % for pdflatex we expect .pdf, .png, or .jpg files
\else%                                 % else we use pure latex
  \usepackage{graphicx}                % allow us to embed graphics files
  \DeclareGraphicsExtensions{.eps}     % for pure latex we expect eps files
\fi%

\graphicspath{{figures/}{pictures/}{images/}{./}} % where to search for the images

\usepackage{microtype}                 % use micro-typography (slightly more compact, better to read)
\PassOptionsToPackage{warn}{textcomp}  % to address font issues with \textrightarrow
\usepackage{textcomp}                  % use better special symbols
\usepackage{mathptmx}                  % use matching math font
\usepackage{times}                     % we use Times as the main font
\usepackage{cite}                      % needed to automatically sort the references
\usepackage{tabu}                      % only used for the table example
\usepackage{booktabs}                  % only used for the table example
\usepackage{amsmath}
\usepackage{bbm}
\usepackage[shortlabels]{enumitem}
\usepackage[ruled]{algorithm2e}
\DeclareMathOperator{\acc}{acc}
\DeclareMathOperator{\conf}{conf}
\DeclareMathOperator{\argmax}{arg\,max}

\newcommand{\edit}[1]{{\color{black} #1}}

\onlineid{1657} %1657

\vgtccategory{Research}
\vgtcpapertype{application/design study}

\title{Calibrate: Interactive Analysis of Probabilistic Model Output}

\author{Peter Xenopoulos, João Rulff, Luis Gustavo Nonato, Brian Barr, Claudio Silva}
\authorfooter{
\item
 Peter Xenopoulos, João Rulff and Claudio Silva are with New York University. E-mail: \{xenopoulos, jlrulff, csilva\}@nyu.edu.
\item
 Luis Gustavo Nonato is with ICMC-USP, São Carlos, Brazil. E-mail:
gnonato@icmc.usp.br.
\item
 Brian Barr is with Capital One. E-mail: brian.barr@capitalone.com.
}

\shortauthortitle{Peter Xenopoulos \MakeLowercase{\textit{et al.}}: Calibrate: Interactive Analysis of Probabilistic Model Output}
\abstract{
Analyzing classification model performance is a crucial task for machine learning practitioners. While practitioners often use count-based metrics derived from confusion matrices, like accuracy, many applications, such as weather prediction, sports betting, or patient risk prediction, rely on a classifier's predicted probabilities rather than predicted labels. In these instances, practitioners are concerned with producing a \textit{calibrated} model, that is, one which outputs probabilities that reflect those of the true distribution. Model calibration is often analyzed visually, through static reliability diagrams, however, the traditional calibration visualization may suffer from a variety of drawbacks due to the strong aggregations it necessitates. Furthermore, count-based approaches are unable to sufficiently analyze model calibration. We present Calibrate, an interactive reliability diagram that addresses the aforementioned issues. Calibrate constructs a reliability diagram that is resistant to drawbacks in traditional approaches, and allows for interactive subgroup analysis and instance-level inspection. We demonstrate the utility of Calibrate through use cases on both real-world and synthetic data. We further validate Calibrate by presenting the results of a think-aloud experiment with data scientists who routinely analyze model calibration.

} % end of abstract


\CCScatlist{ % not used in journal version
 \CCScat{K.6.1}{Management of Computing and Information Systems}%
{Project and People Management}{Life Cycle};
 \CCScat{K.7.m}{The Computing Profession}{Miscellaneous}{Ethics}
}

\teaser{
  \centering
  \includegraphics[scale=0.25]{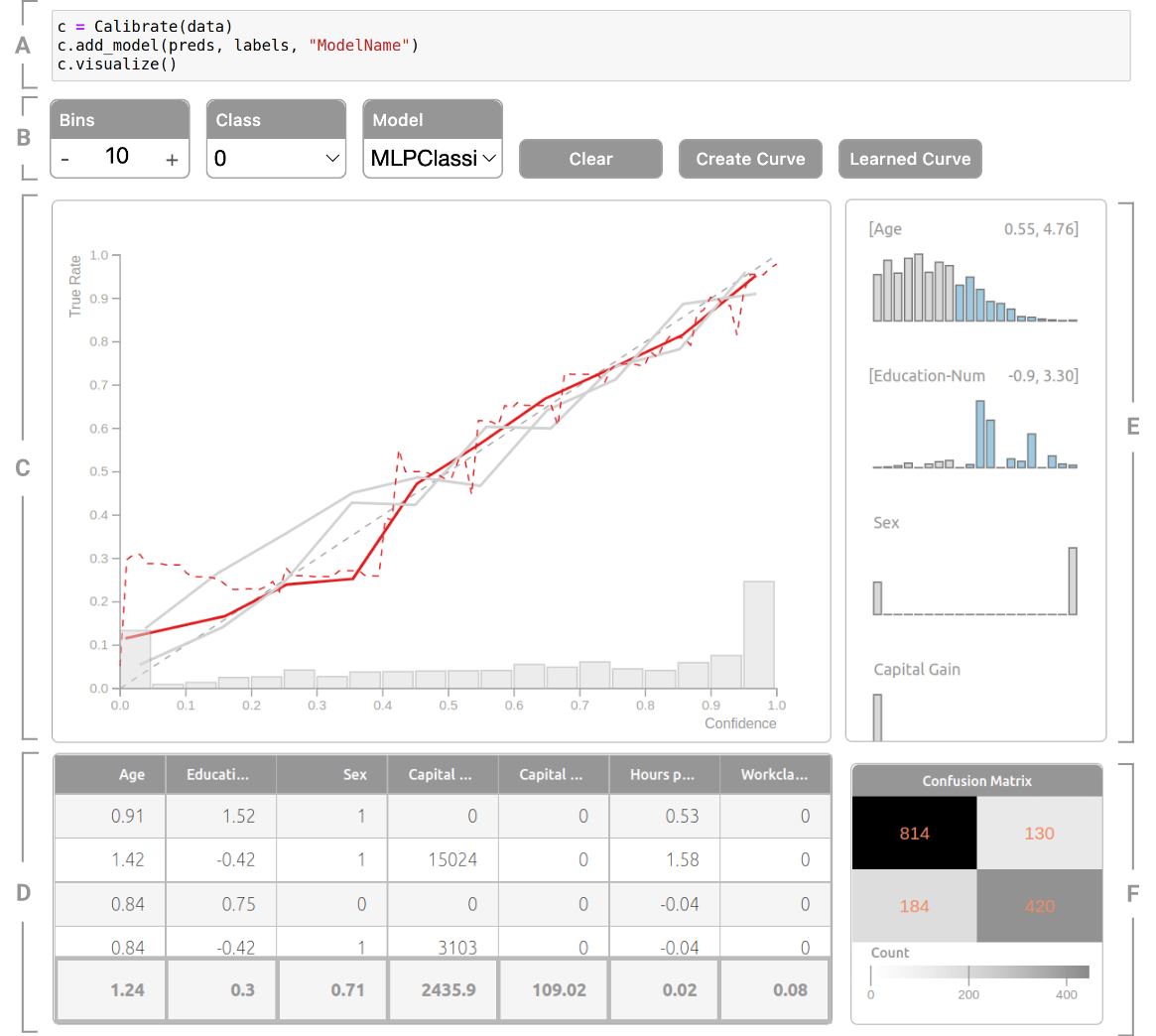}
  \caption{Calibrate visualizing model calibration across different subgroups for income prediction on the Census Income prediction task. A) Calibrate is implemented for easy, one-line use in Jupyter Notebooks. B \& C) Calibration View implements \textit{Learned Reliability Diagrams} (red dotted line), an approach to address shortcomings in traditional reliability diagrams (blue line), as well as a histogram to show the density of predictions. D) Instance View shows instances contained by brushing prediction regions in the Calibration View. E) Feature View allows for subgroup analysis by interactive subgroup creation through brushing feature ranges. F) Performance View shows a confusion matrix for the brushed prediction region.}
  \label{fig:teaser}
}

\vgtcinsertpkg

\begin{document}


\firstsection{Introduction} \label{sec:introduction}

\maketitle

Analyzing classification model performance is an important task for machine learning stakeholders. Furthermore, understanding model performance is a complex task which necessitates an alignment between one's model performance measurements and use case goals. Many machine learning practitioners assess model performance through count-based summary performance metrics, such as accuracy or recall, which rely on predicted labels. The components used to calculate count-based metrics are most commonly viewed in confusion matrices~\cite{townsend1971theoretical}. Such a performance assessment is sensible for situations where practitioners are concerned that their model makes the correct \textit{decisions}. However, when one is interested in a model's predicted \textit{probabilities}, count-based model performance analysis obscures model performance with respect to the practitioner's goals. Model predicted probabilities are important for human decision making, as probabilities are more intuitive for humans than quantities like model scores or log odds~\cite{cosmides1996humans}.

\begin{figure}
    \centering
    \includegraphics[width=\linewidth]{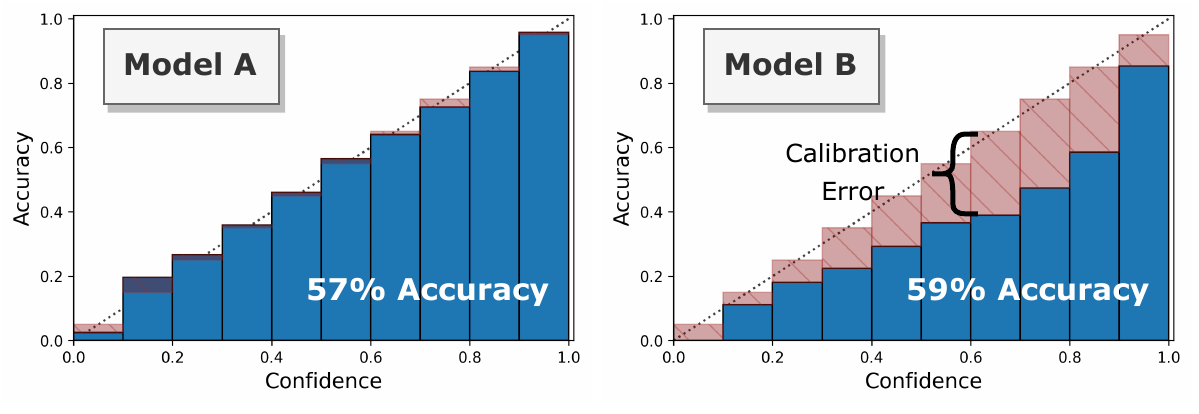}
    \caption{Calibration plots for two deep learning models on the CIFAR100 task. Although Model B attains better accuracy, it systematically achieves a larger calibration error (i.e., the prediction confidence is far from the true prevalence).}
    \label{fig:resnet-example}
\end{figure}

There are many applications which depend on predicted probabilities rather than labels, such as weather forecasting, sports betting, or credit default risk prediction. In these cases, practitioners are especially concerned that their model is \textit{calibrated}. Although there are various formal notions of model calibration, a model is generally considered ``calibrated'' if the model's predicted probabilities correspond to the true class occurrence. Modern machine learning models appear to achieve high performance on count-based metrics, like accuracy, yet can be systematically miscalibrated~\cite{DBLP:conf/icml/GuoPSW17}. For example, consider the two models, based on ResNet-50 and built for the CIFAR-100 task, in Figure~\ref{fig:resnet-example}. While Model B has a slightly higher accuracy, it is much less calibrated than Model A, as there is a large gap between the confidence (predicted probability) and the accuracy (true class rate). 

In addition to count-based metrics, it is also common to assess classifier performance through scoring rules like Brier score or log loss~\cite{brier1950verification, DBLP:books/lib/HastieTF09}. It is known that proper scoring rules are minimized if and only if classifier predictions recover the ground truth conditional distribution for all inputs. However, Kull and Flach show that proper scoring rules, like Brier score and log loss, measure more than \textit{just} calibration~\cite{DBLP:conf/pkdd/KullF15}. Furthermore, calibration is often an important consideration for model fairness. Kleinberg~et~al. show that, if groups have different base rates for their labels, it is statistically impossible to ensure fairness across the balance of the positive and negative classes, as well as the calibration of the model~\cite{DBLP:conf/innovations/KleinbergMR17}. Thus, for many applications, it is critical to deeply understand a classifier's output beyond confusion matrices, count-based metrics, and scoring rule aggregates.

To assess model calibration, practitioners typically turn to visualization, specifically to \textit{reliability diagrams} (shown in Figure~\ref{fig:resnet-example}). Reliability diagrams are a static visualization showing the alignment between the predicted probabilities of the predicted class, referred to as the confidence, and ground-truth class prevalence for the predictions, often referred to as the accuracy. These visualizations require a data transformation that involves a binning stage, whereby predictions are binned by predicted probability, and an aggregation phase, where the average predicted probability and true class prevalence are calculated for each bin. A well-calibrated model will produce results that follow $y=x$, since the model's predictions will reflect the true class rate.

Reliability diagrams form the basis of many calibration-specific model performance metrics, since the binning and aggregation required to create reliability diagrams are identical for many calibration metrics. The aforementioned data transformation requires various hyperparameters, such as the number of bins or the bin creation strategy, which can drastically change the appearance of a reliability diagram~\cite{dimitriadis2021stable}. Furthermore, due to the aggregations that reliability diagrams necessitate, relationships that may be apparent in certain regions of data may be washed out by aggregation. Additionally, the static nature of reliability diagrams hinders subgroup or instance-level analysis. Thus, a visual analytics approach may provide an intuitive link between performance and the underlying data~\cite{DBLP:journals/tvcg/RenALSW17}. 

In this paper, we present Calibrate (Figure~\ref{fig:teaser}), a visual analytics tool to analyze machine learning model calibration. We design and develop Calibrate through interviews with machine learning practitioners who routinely examine model calibration. Calibrate implements Learned Reliability Diagrams, a simple approach to construct reliability diagrams that can capture local calibration relationships. Additionally, Calibrate is integrated with Jupyter Notebooks, which enables analyses to be easily shared and deployed in a wide array of machine learning environments. We make the following contributions:

\begin{itemize}
    \item Learned Reliability Diagrams, a new approach to construct reliability diagrams resistant to the pitfalls of conventional reliability diagrams.
    \item Calibrate, an interactive visual analytics tool to analyze model calibration. We design Calibrate to fulfill requirements derived from interviews with machine learning practitioners. Calibrate is designed with Jupyter Notebooks use in mind.
    \item An evaluation of Calibrate through use cases and expert interviews showing how Calibrate can be used to analyze calibration across subgroups and how we can use Calibrate to understand determinants of model calibration. 
\end{itemize}

% The rest of the paper is structured as follows. In Section~\ref{sec:related-work}, we review related work in calibration and model performance analysis. Section~\ref{sec:calibration-bg} provides a background on calibration including definitions of different notions of calibration, a formal introduction to reliability diagrams, and a review of reliability diagram-derived calibration metrics. Section~\ref{sec:calibrate} introduces the design requirements of Calibrate, which we develop from interviews with machine learning practitioners, and explains the functionality and implementation of the system. We also describe our new approach to construct reliability diagrams as well as provide uncertainty estimates for these diagrams. In Section~\ref{sec:evaluation} we perform an evaluation of Calibrate through case studies and expert interviews. Sections~\ref{sec:discussion} and \ref{sec:conclusion} discuss our results and conclude the paper, respectively. 
\section{Related Work} \label{sec:related-work}
\subsection{Model Calibration}
Calibration is a long-studied topic, particularly from the statistics, metereologic, and more recently, machine learning communities. However, there is limited visualization-specific work regarding calibration. In 1920, Hallenbeck introduced a tabular format to assess the probabilistic forecast of rain~\cite{hallenbeck1920forecasting}. Reliability diagrams are a direct mapping from the tabular format introduced by Hallenbeck to a line plot, and are a standard calibration visualization~\cite{sanders1963subjective,murphy1977reliability,lichtenstein1977calibration,degroot1983comparison}. Reliability diagrams effectively bin predictions based on an observation's predicted probability and compares the average classifier probability prediction of the bin to the average rate of the true label in that bin. Hagedorn~et~al. encodes the size of each bin through each marker's corresponding size in the reliability diagram~\cite{hagedorn2005rationale}. Br\"{o}cker and Smith develop consistency bars to alleviate the problem of uneven distributions in bins~\cite{brocker2007increasing}. More recently, Vaicenavicius~et~al. introduce a method to visualize calibration for three- and four-class problems~\cite{DBLP:conf/aistats/VaicenaviciusWA19}.

Most prior work in calibration has revolved around metrics to assess a model's calibration or methods to calibrate a model. Brier first developed the Brier score in the 1950s to assess weather forecasts~\cite{brier1950verification}. Brier score, like log loss, is a proper scoring rule, meaning that it is minimized when the probability distribution output by the classifier recovers the true conditional distribution~\cite{winkler1969scoring, DBLP:books/lib/HastieTF09}. Kull and Flach show that proper scoring rules can be decomposed into a sum of various losses, one of which is calibration loss. Thus, proper scoring rules do not \textit{only} capture calibration loss. 

Guo~et~al. demonstrate that modern neural architectures often produce uncalibrated outputs~\cite{DBLP:conf/icml/GuoPSW17}. Furthermore, they find that miscalibration can worsen even as classification error is reduced. To measure the calibration loss, they use a variety of metrics, such as expected and maximum calibration error (ECE and MCE, respectively), which measure calibration error relative to a reliability diagram~\cite{DBLP:conf/aaai/NaeiniCH15}. Nixon~et~al. propose a new metric, adaptive calibration error, and show improvement over standard calibration metrics~\cite{DBLP:conf/cvpr/NixonDZJT19}. Vaicenavicius~et~al. propose a hypothesis test for the aforementioned calibration metrics to provide a more rigorous analysis of model calibration. Widmann~et~al. generalize common calibration metrics, like ECE and MCE, for multi-class problems~\cite{DBLP:conf/nips/WidmannLZ19}. \edit{While there has been much work done towards \textit{quantifying} model calibration characteristics, there has been little work directed towards \textit{visualizing} model calibration.}

\subsection{Assessing Model Performance}
The ability to visualize common model performance metrics, like accuracy, precision, and recall, is a fundamental task for machine learning practitioners. Thus, the visualization capabilities needed to analyze model performance metrics are common to most machine learning platforms and libraries~\cite{DBLP:journals/tvcg/HohmanKPC19}. For example, the \texttt{sklearn} Python library allows one to easily visualize confusion matrices, ROC curves and reliability diagrams. Likewise, TensorFlow allows a user to visualize relevant performance metrics model comparison~\cite{DBLP:journals/corr/AbadiABBCCCDDDG16}. Hinterreiter~et~al. identify three levels of detail for model analysis, namely global-, class- or instance-level~\cite{DBLP:journals/tvcg/HinterreiterRSE22}. Global-level considers classifier performance across the whole data set, and may consider global metrics like accuracy. Class-level summarizes a model's performance over specific classes in the data. In practice, this may look like class-specific versions of global metrics or confusion matrices. Instance-level focuses on the errors of specific observations, based on predicted labels or probabilities.

Much class-level work revolves around the information contained within or derived from confusion matrices, a fundamental visualization in model performance analysis. Alsallakh et al. introduce the confusion wheel, which visualizes confusion between classes for a single classifier~\cite{DBLP:journals/tvcg/AlsallakhHHMR14}. Gleicher~et~al. present Boxer, a visual system to explore the results of multiple classifiers~\cite{DBLP:journals/cgf/GleicherBYH20}. Recently, G{\"o}rtler~et~al.~ detail Neo, a system that generalizes confusion matrices~\cite{gortler2021neo}. Similar to class-level analysis, there is also strong demand from machine learning practitioners to analyze their models across subgroups in their data. Cabrera~et~al.~propose FairVis, a visual analytics system which uses multiple coordinated views to investigate fairness metrics across subgroups of interest~\cite{DBLP:conf/ieeevast/CabreraEHKMC19}. Dingen~et~al.~introduce RegressionExplorer, a visual analytics tool which enables users to compare logistic regression models across subpopulations~\cite{DBLP:journals/tvcg/DingenVHMKBW19}. However, the aforementioned approaches do not allow for instance-level.

Increasingly, many class-level visualizations are turning towards designs that also allow for instance-level inspection. Amershi et al. propose ModelTracker, which arranges observations as boxes on a one-dimensional axis that conveys prediction score~\cite{DBLP:conf/chi/AmershiCDLSS15}. These boxes allow a user to select instances of interest, while grouping instances in an easy-to-understand way. Ren et al. describe Squares, an interactive system that also visualizes observations using boxes, and is built for multiclass classification problems~\cite{DBLP:journals/tvcg/RenALSW17}. Kahng~et~al. propose ActiVis, an interactive visualization system that integrates several coordinated views to explore deep learning models~\cite{DBLP:journals/tvcg/KahngAKC18}. \edit{Although the aforementioned approaches may allow for subgroup analysis or instance-level inspection, none of them directly analyze model calibration}. 

% While some of the above systems order instances by their predicted probability or their logarithmic loss, none of the above systems directly addresses model calibration.

% workflow for visual diagnostic of binary classifiers (class-level, instance level)~\cite{DBLP:conf/ieeevast/KrauseDSAB17}

% error discovery through semantic data exploration (anchorviz, instance-level?)~\cite{DBLP:conf/iui/ChenSVRDS18}

% interactive comparison of classifier results (subset, class level)~\cite{DBLP:journals/cgf/GleicherBYH20}

% https://arxiv.org/pdf/2110.00530.pdf fairness data sets
\section{Calibration Background} \label{sec:calibration-bg}
In this section, we provide a background on calibration. First, we enumerate and define various notions of calibration in Section~\ref{sec:notions-calibration}. Then, in Section~\ref{sec:reliability-diagrams}, we introduce reliability diagrams. Next, in Section~\ref{sec:calibration-metrics}, we define common calibration metrics. Finally, we outline issues with reliability diagrams in Section~\ref{sec:reliability-diagram-issues}. We denote our classifier as $\widehat{\textbf{p}}: X \rightarrow Y$, which outputs class probabilities for $1, ..., K$ classes. Any instance $\textbf{x} \in X$ input to $\widehat{\textbf{p}}$ outputs a probability vector $\widehat{\textbf{p}}(\textbf{x})$.

% We denote $X$ as our input space, $Y$ as our output space and $f: X \rightarrow Y$ as our classifier. We assume the classification problem has $K$ total classes.

% topology for reliability, color nodes by discrepancy of predicted to actual (calibration error)
% generate lime or shap counterfactual explanations
% for use cases where units of explanation are very important, then calibration
% don't need to do calibration for a single model. could be a system of models (give an example)

\subsection{Notions of Calibration} \label{sec:notions-calibration}
Filho~et~al. outline several types of model calibration~\cite{DBLP:journals/corr/abs-2112-10327}. The first, and most common, type of calibration, coined \textit{confidence calibration}, requires that among all instances where the probability of the most likely class is predicted to be $\alpha$, the expected accuracy is also $\alpha$, where $\alpha \in [0,1]$. This is a commonly deployed notion of calibration since it is easy to implement, as one only has to consider the highest predicted probability. Thus, a classifier is confidence calibrated if

\begin{equation}
    \mathbbm{P}(Y = \argmax (\widehat{\textbf{p}}(X)) \mid \max (\widehat{\textbf{p}}(X)) = \alpha) = \alpha 
\end{equation}

%\begin{equation}
%    \mathbbm{P}(Y = \widehat{Y} \mid S_{\widehat{Y}} = \alpha) = \alpha \textrm{ where } \widehat{Y} = \argmax_{i} S_i
%\end{equation}

% \noindent where $S_i$ denotes the $i$-th dimension of the output $f(x)$, where $x \in X$.

Another form of calibration is \textit{classwise calibration}, proposed by Zadrozny and Elkan~\cite{DBLP:conf/kdd/ZadroznyE02}. Classwise calibration requires that all one-vs-rest probability estimators, derived from the original model $\widehat{\textbf{p}}$, are calibrated. That is, for class $i \in {1, ..., K}$ and a predicted probability $s$, 

\begin{equation}
    \mathbbm{P} (Y = i \mid \widehat{p_i}(\textbf{x}) = s) = s
\end{equation}

\noindent where $\widehat{p_i}(\textbf{x})$ represents the $i$-th index of $\widehat{\textbf{p}}(\textbf{x})$.

Finally, the strongest notion of calibration is \textit{multiclass calibration}. Given prediction vector $\textbf{q} = (q_1, ..., q_k) \in Y$, $\widehat{\textbf{p}}$ is multiclass calibrated if the proportion of classes among all possible instances on $\textbf{x} \in X$ getting the same prediction $\widehat{\textbf{p}}(X) = \textbf{q}$ is equal to the prediction vector $\textbf{q}$. 

\begin{equation}
    \mathbbm{P} (Y = i \mid \widehat{\textbf{p}}(\textbf{x}) = \textbf{q}) = q_i \textrm{ for } i \in 1,...,K
\end{equation}

\subsection{Reliability Diagrams} \label{sec:reliability-diagrams}
Reliability diagrams are a popular method to assess model calibration~\cite{degroot1983comparison, DBLP:conf/icml/Niculescu-MizilC05}. Reliability diagrams group predictions into discrete bins and 
plot the expected accuracy on the y-axis and average classifier confidence on the y-axis. We call this line calibration curve. A calibrated model is one where the difference between the expected accuracy and the average classifier confidence is small. Thus, a perfectly calibrated model would follow the line $y = x$. Reliability diagrams are specified for a particular class of interest.

To estimate the expected accuracy, we group data into ``bins''. Typically, this is done by dividing predictions into $W$ bins of width $1/W$ along the range of $[0,1]$. We let $B_w$ be the set of indices of observations whose prediction confidence is contained within the interval $I_w = (\frac{w-1}{W}, \frac{w}{W}]$. The accuracy of $B_w$, for a given class $i$, is defined as

\begin{equation}
    \acc(B_w) = \frac{1}{|B_w|} \sum_{j \in B_w} \mathbbm{1}(\widehat{y}_j = y_j)
\end{equation}

\noindent where $\widehat{y}_j$ and $y_j$ are the predicted and true class labels for observation $\textbf{x}_j$, and $\mathbbm{1}$ is an indicator function. We can define the confidence of $B_w$ as 

\begin{equation}
    \conf(B_w) = \frac{1}{|B_w|} \sum_{j \in B_w} \widehat{p}_i (\textbf{x}_j)
\end{equation}

\noindent where $\widehat{p}_i$ is the predicted probability of class $i$ for observation $j$.

\begin{figure*}
    \centering
    \includegraphics[width=\linewidth]{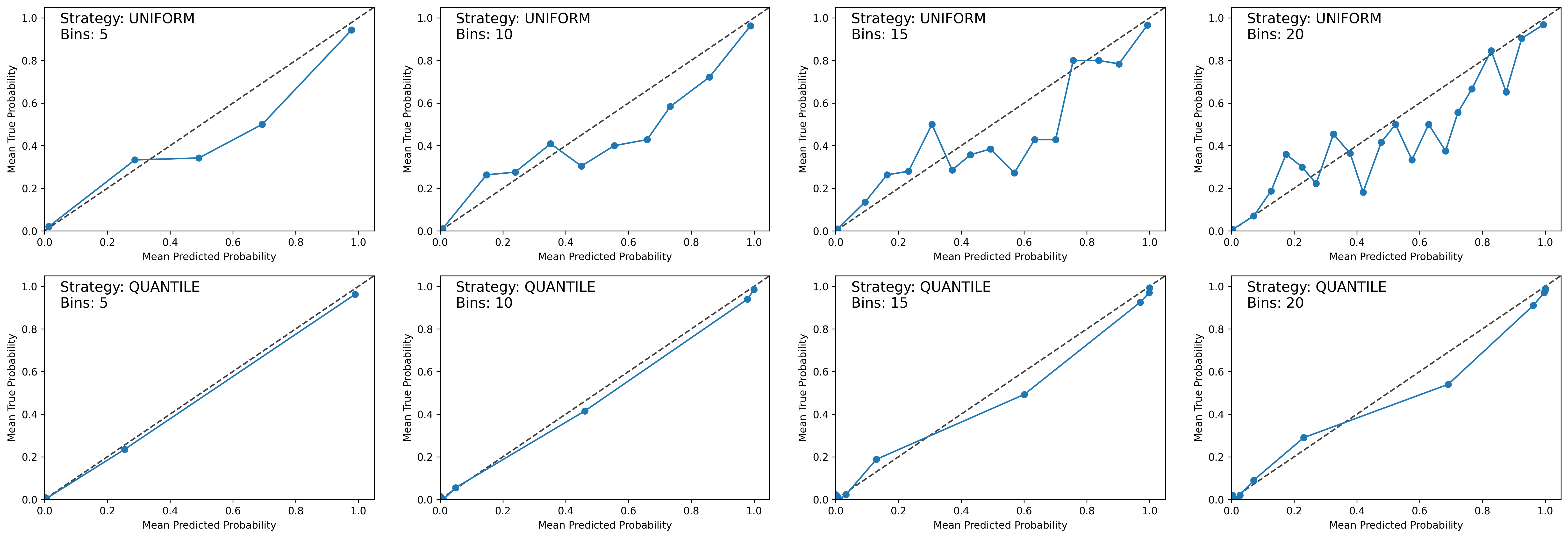}
    \caption{Hyperparameters of reliability diagrams, namely the number of bins and binning strategy, can not only drastically change a diagram's visual representation but also the calculation of downstream metrics like Expected or Maximum Calibration Error. For example, the ``quantile'' strategy (bottom row), which creates bins of equal number of observations, would suggest the above model is well-calibrated. However, the ``uniform'' strategy (top row), which creates bins of identical width, suggests that the model is not well-calibrated between predicted probabilities of 0.5 and 0.9.}
    \label{fig:reliability-diagram}
\end{figure*}

In some implementations, like \texttt{sklearn}, users define the number of bins, and then a strategy which will either generate bins of equal number of samples or of equal width. These parameters can be difficult to select, as they have drastic implications on the reliability diagram itself. We show an example of a reliability diagram in Figure~\ref{fig:reliability-diagram}, as well as demonstrate how the parameters, namely the number of bins and binning strategy, can drastically change the visual representation of model calibration.

\subsection{Calibration Metrics} \label{sec:calibration-metrics}

Proper scoring rules are loss measures that are minimized when the predicted class distribution equals the true class posterior distribution~\cite{DBLP:conf/pkdd/KullF15}. The two most well-known proper scoring rules, which are also used as surrogate losses for optimization, are Brier score and log loss. Brier score ($\phi_{BS}$) and log loss ($\phi_{LL})$ are defined as 

\begin{equation}
\begin{split}
\phi_{BS} &= \frac{1}{N} \sum_{i=1}^N \sum_{j=1}^K (\widehat{y}_{ij} - y_{ij})^2 \\
\phi_{LL} &= - \frac{1}{N} \sum_{i=1}^N \sum_{j=1}^K y_{ij} \cdot \log(\widehat{y}_{ij}) \\
\end{split}
\end{equation}

\noindent where $N$ is the total number of observations, $K$ is the total number of classes, $\widehat{y}_{ij}$ is the predicted probability of class $i$ for observation $j$, and $y_{ij}$ is equal to $1$ if $i$ is the true class of observation $j$ and $0$ otherwise.

While scoring rules are minimized when the predicted class distribution from the classifier equals the true class posterior, we also know that proper scoring rules measure more than just calibration loss. This decomposition is shown in~\cite{DBLP:conf/pkdd/KullF15}. Thus, we can construct strict measures of calibration using the information calculated that we use in reliability diagrams, namely, the accuracy and confidence of each bins. From these quantities, we can calculate the expected calibration error (ECE) and maximum calibration error (MCE)~\cite{DBLP:conf/aaai/NaeiniCH15}. We define ECE and MCE as

\begin{equation}
\begin{split}
ECE &= \sum_{w=1}^W \frac{|B_w|}{N} | \acc(B_w) - \conf(B_w) | \\
MCE &= \max_{w \in \{1,...,W\}} | \acc(B_w) - \conf(B_w) |
\end{split}
\end{equation}

% While there has been recent work in proposing new calibration metrics, ECE and MCE are the most common~\cite{DBLP:conf/cvpr/NixonDZJT19}. 

\subsection{Issues with Reliability Diagrams} \label{sec:reliability-diagram-issues}

\edit { Reliability diagrams are simple to compute, requiring just a single pass through a model's predictions. In this pass, one assigns each prediction to a bin. Each bin contains a range of predictions, like those of $[0,0.1)$, $[0.1, 0.2)$, and so on. The aforementioned simplicity makes sense from a historical point of view -- Hallenbeck tabulated rain forecasts into a table of ten bins in 1920, long before modern computing~\cite{hallenbeck1920forecasting}. However, we still see influences of Hallenbeck's design choices in modern day calibration analysis. For example, many works still use ten bins or fewer~\cite{DBLP:conf/icml/GuoPSW17, DBLP:conf/uai/Niculescu-MizilC05, DBLP:conf/icml/ZadroznyE01, DBLP:conf/icml/Niculescu-MizilC05}. The number of bins is important since it strongly influences the construction of popular calibration metrics, such as expected calibration error. Nixon~et~al.~detail various problems with expected calibration error due to the design choices stemming from the binning and aggregation procedures required for a reliability diagram~\cite{DBLP:conf/cvpr/NixonDZJT19}. In turn, these issues may also be associated with reliability diagrams. We discuss a few salient issues below. 

\label{issue:1}\textbf{Issue 1: Variability from Bin Selection.} Even small changes for the number of bins can have drastic effects on the produced reliability diagram. Furthermore, binning strategies, such as fixed bin width or quantile binning may induce certain shapes on the reliability diagram depending on the distribution of the predictions. Effectively, a large bin count may create many small bins that have high variance. On the other hand, a small bin count may wash out local calibration information.

\label{issue:2}\textbf{Issue 2: Competing Bin Effects}. One issue with reliability diagrams is that binning may wash out competing effects between overconfident and underconfident predictions in a bin. In practice, such an occurrence is common, and it hinders the ability of a practitioner to identify regions of interest in a reliability diagram. In the worst case, the effects would cancel to show no calibration error for a given bin, even though there may be substantial miscalibration within that bin. Thus, the user-specified binning parameters may have strong impacts. 

% \edit{This is a similar issue to the number of bins used to display distributions in histograms, and Sahann~et~al. perform a user study on this issue~\cite{DBLP:conf/visualization/SahannMS21}. They find that the visual perception of the underlying data depended on the number of bins, yet for a certain number of bins, adding further bins does not help improve a user's perception of the underlying distribution. We denote issues regarding binning parameters as \textbf{S1}\label{issue:binning}.}

\label{issue:3}\textbf{Issue 3: Importance of Prediction Distribution}. The distribution of predicted class probabilities is unlikely to be uniform in practice. In such a case, just a few bins will contribute the most to the expected calibration error. However, in modern reliability diagrams, each mark representing a bin is typically a single point; bin size is not usually visually encoded. Thus, it is possible for two reliability diagrams to look identical, yet have vastly different calibration errors due to the underlying distribution of predicted probabilities. The issue of conveying bin sample size is also important for uncertainty estimation -- bins with low counts of observations will inherently have larger confidence intervals for their estimated average class prevalence. Possible solutions include encoding bin size through marker's size~\cite{hagedorn2005rationale} or by computing a confidence interval to visualize on the reliability diagram itself~\cite{brocker2007increasing}. } 
\section{Calibrate} \label{sec:calibrate}
%Lastly, each practitioner's workflow used very little interaction, relying mostly on the interactivity provided by Jupyter Notebooks with creating, running and deleting code cells. 
In this section, we describe Calibrate, a visual analytics tool that enables interactive analysis of model calibration. First, we present the results of interviews with machine learning experts who routinely examine model calibration. We use these interviews to construct a list of design requirements. Then, we outline the various views in Calibrate, and how these views address the requirements identified in Section~\ref{sec:design-goals}. Finally, we discuss the implementation of Calibrate, which is available for use in Jupyter notebooks. We show an example of Calibrate applied to a predictive model trained with real world data, along with a brief description of Calibrate's views, in Figure~\ref{fig:teaser}.

\subsection{Design Requirements} \label{sec:design-goals}
We conducted interviews with four machine learning practitioners who actively work on developing, refining and deploying machine learning models, especially those which predict probabilities. We denote each practitioner as P1 through P4. Each practitioner was familiar with reliability diagrams and used them in their work. We asked each expert about their usage of reliability diagrams, particularly how they create them, what types of tasks they attempt to complete using reliability diagrams, and what issues they have encountered using reliability diagrams in practice. Where relevant, we link expert interview feedback to issues described in Section~\ref{sec:reliability-diagram-issues}.

% State how common reliability diagrams are, how they are used
For each practitioner, reliability diagrams were a common tool used in their analyses. P2 remarked that, especially recently, their organization is using more reliability diagrams to evaluate their models. A common characteristic among all practitioners was that reliability diagrams were most used in content created for other technical staff, such as analysts, data scientists, or data engineers, rather than for less technical stakeholders. However, P4 also mentioned that reliability diagrams may be more intuitive when they present a model's performance to non-technical stakeholders, noting that reliability diagrams are ``generally easy to explain'' and they ``answer a natural question.'' Most practitioners engaged in binary classification tasks. For multiclass problems, they would typically use a one-versus-rest approach when constructing reliability diagrams. 

% subset analysis as a major task
All practitioners identified subset analysis as a routine task when analyzing model calibration with reliability diagrams. This task involved manually filtering data to create subgroup of interest and then creating reliability diagrams for these subgroups. Each practitioner mentioned that subgroup analysis was primarily a manual process and involved much domain knowledge to construct subgroups. P1 mentioned that typically they went into this subset analysis knowing specific subsets to analyze, based on domain knowledge. On the other hand, P3 remarked that they would ``sanity check'' their model across broad subgroups, particularly across subgroups that they may expect to be harder to predict. While these subgroups would be defined by filtering the data by its features, P4 noted that sometimes, although rarely, they would construct subgroups by manually grouping instances.

% how to fix subsets
When practitioners found a subgroup which was miscalibrated, many of them turned to feature engineering rather than post-hoc corrections like Platt scaling or isotonic regression. For example, when they found a miscalibrated subgroup, P4 mentioned they would typically go back and attempt to create new features to target that specific subgroup. Similarly, P1 remarked that when they found an important subgroup which was miscalibrated, they would either perform more feature engineering, or they would create a separate model for their subgroup. In some cases, practitioners would change their model choice, as there was some understanding among the practitioners that model architecture had an impact on calibration. There were also shared concerns on class imbalance and their effects on calibration by P1, P2 and P4. We investigate this topic further in Section~\ref{sec:determinants-use-case}.

% distributions
Displaying the density of the predictions was a common complement to reliability diagrams. In particular, this visualization was deemed by most practitioners to be important for understanding where their model's calibration estimate was more unreliable \edit{(Issue 3)}. Intuitively, where one has lower density of predictions, one would expect the ``error'' to be higher, as mentioned by P2. P1 mentioned that they sometimes used error bars on their reliability diagrams, but oftentimes opted to plot the distribution of the predictions underneath the reliability diagram, as most other stakeholders could infer which regions had highest variability. Furthermore, P1 described uncertainty estimation in common calibration libraries as having little support.

% implementation, how do they achieve this task?
A common issue with conventional reliability diagrams was selecting the number of bins. Most practitioners used the default parameters or between 10 and 20 bins. P1 mentioned that they had ``only ever seen bins between 10 to 20'', and that they found a many bins as unhelpful in analyzing model calibration. P4 noted that they were not familiar with ``industry practice'' and that they had concerns over reliability diagrams potentially returning different conclusions due to different numbers of bins \edit{(Issue 1)}. Although the practitioners were seemingly less worried about competing effects within bins, P1 mentioned that ``the binning obscures a lot of fine detail'' \edit{(Issue 2)}, and that the static nature of reliability diagrams does not encourage exploration.

% implementation, deployment
The practitioners unanimously stated that they implement reliability diagrams using popular packages in Python and R, and in particular, in Jupyter Notebooks. For example, P1 mentioned that they ``perform virtually all model analysis and produce all reliability diagrams in Jupyter''. One common aspect mentioned by the practitioners was that Jupyter Notebooks are easily reproducible, and therefore they were a desired for disseminating model performance analysis to other stakeholders. Additionally, P2 mentioned that much of their organization's model performance analysis is done in notebooks where the model building also takes place. P3 noted that although model training occurred outside of Jupyter, they would frequently read model predictions into a Jupyter notebook where they would analyze the model's calibration. 

From the above interviews, we compiled the following requirements:

\begin{enumerate}[start=1,label={\bfseries R\arabic*}]
% guide users towards areas on interest
\item \label{req:lrd} \textbf{Allow for identification of interesting calibration regions.} Hyperparameter choice is a difficult task for our practitioners and indeed has effects on the visual representation of model calibration. In many cases, practitioners simply use the default parameters given by whichever library they use. However, as we describe in Section~\ref{sec:calibration-bg}, traditional diagrams suffer from a variety of issues. Our system should address the issues with traditional reliability diagrams, while also allowing a user to easily change the parameters for conventional reliability diagrams.

% instance review
\item \label{req:instance} \textbf{Connect performance to data, especially in bins.} Due to the static nature of reliability diagrams, it is difficult for practitioners to inspect specific bins and analyze the instances that comprise the selected bin. Users should be able to select predictions regions, aside from the bins returned by conventional reliability diagrams, and analyze the instances contained within the region.

\item \label{req:subgroup} \textbf{Allow for subgroup analysis for reliability diagrams.} A fundamental task is to analyze calibration on subgroups of predictions. Finding these subgroups is generally a manual process, whereby the practitioners rely on prior knowledge to manually define and filter subgroups. Thus, we should allow for users to interactively investigate and compare calibration among subgroups of predictions.

% These findings would then be used to iterate on the model building process by hypothesizing about why particular subgroups perform poorly or by modifying the feature engineering process to target improved performance in subgroups.
% , and to propose potentially interesting subgroups to the user

\item \label{req:jupyter} \textbf{Integrate with Jupyter Notebooks.} The practitioners that we interviewed made their reliability diagrams through calibration-specific packages in Python and to a lesser degree, R. Overwhelmingly, they also deployed the code to produce these diagrams in Jupyter Notebooks, which they found easier to disseminate to other stakeholders in their analytics processes. Furthermore, because many practitioners built their models in Jupyter, they prefer to perform their model analysis near to the model training workload. Therefore, it is important for the tool's implementation to be compatible with Jupyter Notebooks.
\end{enumerate}

% https://arxiv.org/pdf/1904.01685.pdf
% talk about RDs, uncertainty, loop back to interviews
\subsection{Learned Reliability Diagrams}
Reliability diagrams are effectively an attempt to estimate the relationship between a model's predicted probabilities and their associated true outcomes. However, since we only know an observation's label outcomerather than its true probability, we use binning to estimate the ``true probability'' for a collection of samples. In practice, traditional approaches to reliability diagrams use 10 to 20 bins~\cite{DBLP:conf/icml/GuoPSW17}. Typically, these bins do not overlap, and they encompass distinct, uniform ranges of the model predictions. One variation is to use a ``quantile'' binning strategy, whereby bins are created according to the distribution of the data. However, as we show in Figure~\ref{fig:reliability-diagram}, the quantile binning strategy, for some model outputs, can cause most bins to be created near 0 or 1, which complicates analysis of model calibration between the extreme predictions. Another approach to solve the aforementioned issues with reliability diagrams is to use overlapping bins. To the best of our knowledge, the only prior work using overlapping bins is that of Caruana and Niculescu-Mizil, which sorts predictions and creates rolling bins of 100 samples each to calculate confidence and accuracy~\cite{DBLP:conf/kdd/CaruanaN04}. One issue with overlapping bins is that the distribution of the data may result in bins that encompass large ranges of the predicted probability distribution. Thus, local calibration changes may be muted. 

% Traditional approaches to reliability diagrams use 10 to 20 bins, in practice~\cite{DBLP:conf/icml/GuoPSW17}. Typically, these bins do not overlap as they encompasses distinct ranges of the model predictions. However, to \textit{visualize} calibration, it is not strictly necessary for one to consider exclusive bins. To our knowledge, the only prior work using overlapping bins is that from Caruana and Niculescu-Mizil, which sorts predictions, then creates rolling bins of 100 samples each and calculates their positive class prevalence and the average predicted probability of the positive class~\cite{DBLP:conf/kdd/CaruanaN04}. One issue of the aforementioned approach is that the distribution of the data may result in bins that encompass large ranges of the predicted probability distribution. Thus, local calibration changes may be muted.

\begin{figure}
    \centering
    \includegraphics[width=\linewidth]{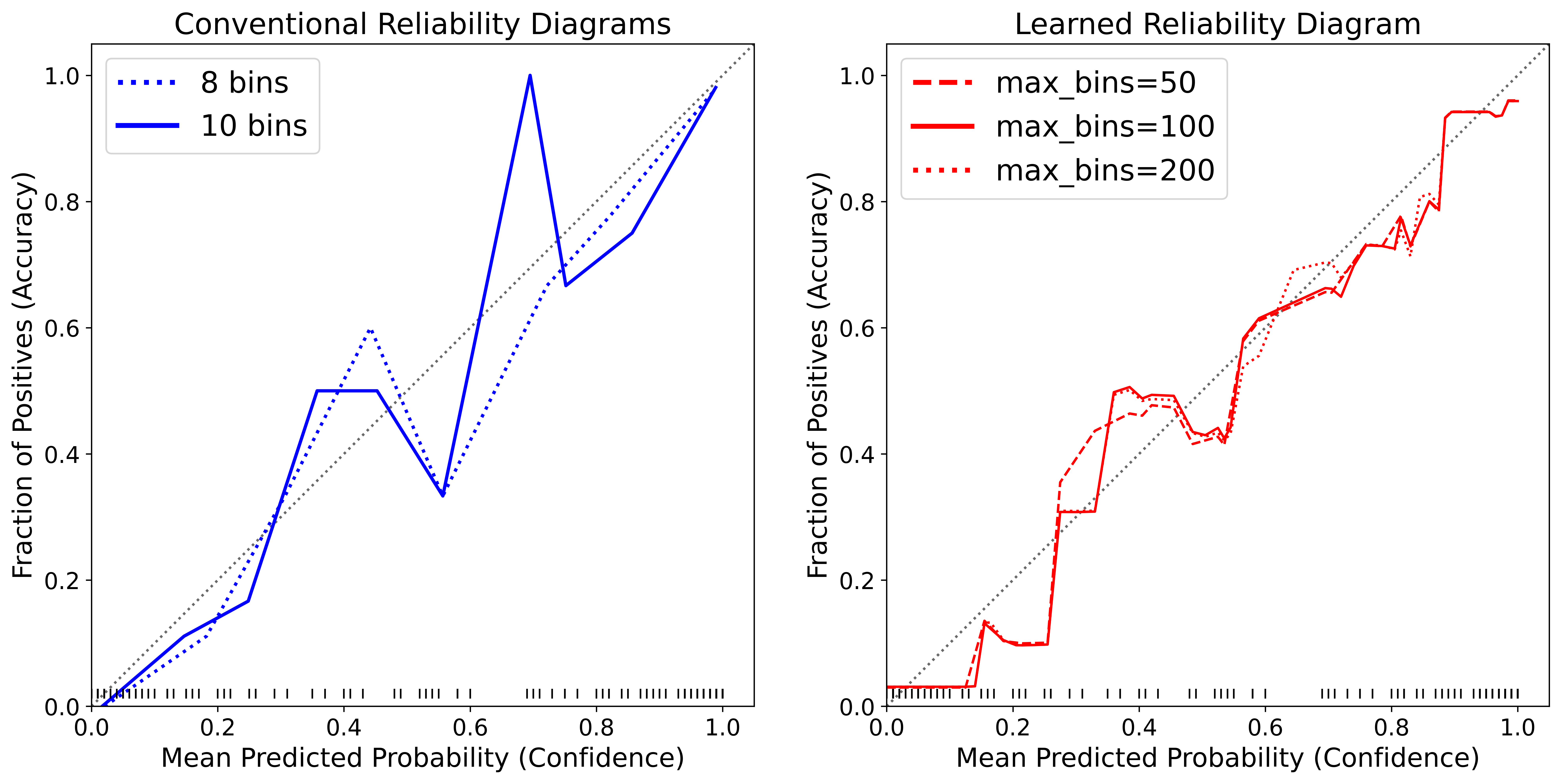} % figures/lrd_wide.png _equal
    \caption{ \edit{Example of conventional 8 and 10 bin diagrams (left) along with learned reliability diagrams for a random forest classifier trained on the Wisconsin Breast Cancer dataset. The density of the predictions is shown below each graph using a rug plot. Learned reliability diagrams are resistant to parameter change, while simply going from 8 to 10 bins induces large changes.} }
    \label{fig:lrd}
\end{figure}

\edit{In general, setting the appropriate binning parameters, such as the number of bins or binning strategy, is a difficult activity for many practitioners, and many rely on comfortable past choices. The choice of bins can lead to drastically different conclusions. In Figure~\ref{fig:lrd}, we show various bin sizes for predictions from a random forest classifier trained on the Wisconsin Breast Cancer dataset. By simply considering ten instead of eight bins, we see a large difference in the conclusion one may draw about the model's calibration. Oftentimes, a small number of bins may suggest the model is calibrated, as the bins encompass larger regions, and a large number of bins suggests the opposite, as the bins have few samples in them.}

\edit{To address the issues arising from binning parameter selection for conventional reliability diagrams (\ref{req:lrd}), we propose to \textit{learn} reliability diagrams. Taking a set of predicted probabilities $X$, along with the true labels for these predictions $Y$, we learn a new function $f: X \rightarrow Y$. This univariate function can be learned through any classifier that produces probabilistic predictions, such as a boosted tree or a generalized additive model. If our underlying predictions are calibrated, then $f$ will closely follow the 45-degree line for all inputs. Our approach is similar to post-hoc calibration. In post-hoc calibration, one attempts to fix a model's predicted probabilities after the model has generated its predictions. Typically, one imposes a monotonic functional form on $f$, such as through a sigmoid (known as Platt scaling)~\cite{platt1999probabilistic} or isotonic regression~\cite{DBLP:conf/kdd/ZadroznyE02}. However, in our case, the goal is not to fix the original predicted probabilities, but rather \textit{summarize} their relationship with the true probabilities in a manner that is less sensitive to binning parameter choices. Thus, while in practice $f$ can be any classifier, we generally want to choose $f$ in a way that is agnostic to the $f$'s shape (e.g., using logistic regression for $f$ would impose a sigmoid shape) and in a manner that is stable under parameter changes, since many of choices of $f$ would require specific parameters. After learning $f$, we can plot $f(x)$ for all $x \in [0,1]$ to visualize the \textit{learned reliability diagram} (Figure~\ref{fig:lrd}). A learned reliability diagram is understood the same way as a conventional reliability diagram. We can still calculate ECE for our learned reliability diagram by determining the total area between the curve and the 45-degree line. Lastly, since our prediction task only takes a single variable as input (the original model predictions), learned reliability diagrams are fast to generate.}

\edit{By estimating the relationship between the predicted and true probabilities through a continuous function rather than through discrete intervals, we gain a few advantages. First, the issues of competing effects within a bin or variability imposed by binning parameter selection are largely avoided since learned reliability diagrams are beholden to the distribution of the predictions rather than arbitrary, user-selected binning parameters. Furthermore, many choices of $f$ grant us the ability to easily extract confidence interval estimates. For our work, we use Explainable Boosting Machines (EBMs)~\cite{DBLP:conf/kdd/LouCG12, DBLP:conf/kdd/LouCGH13, DBLP:journals/corr/nori-interpret}, which are available in the \texttt{interpret} Python package. While some choices of $f$ may still necessitate parameter tuning, we find that EBMs provide stable results across a wide parameter set. For example, in Figure~\ref{fig:lrd}, we vary the ``max bins'' parameter in EBMs and find that the resulting learned reliability diagrams are quite resistant to parameter changes.}

%\begin{figure}
%    \centering
%    \includegraphics[width=\linewidth]{figures/learned_reliability_diagrams.pdf} % 
%    \caption{Calibration results from a model trained on a synthetic data set. A) Reliability diagrams using a traditional 10 bin approach, Explainable Boosting Machines, and splines. We use a rug plot to show the distribution of the predictions. B) 95\% confidence intervals derived from spline GAM. C) Explainable boosting machines learned diagrams by bin parameter choice. D) Spline learned diagrams by number of splines parameter choice.}
%    \label{fig:lrd}
%\end{figure}

\subsection{Calibration View}
Building upon users' familiarity with reliability diagrams, the Calibration View (Figure~\ref{fig:teaser}-C) visualizes both traditional and learned reliability diagrams (\ref{req:lrd}). Users may change the number of bins through a selection box at the top of the view (Figure~\ref{fig:teaser}-B). To visualize calibration for different classes, we allow a user to select the class of interest, which presents the confidence calibration for the selected class. Additionally, users may interact to display learned reliability diagrams or to clear the current view. The calibration view is tied to the other views through select and brushing operations. Since a user may create many calibration curves, we allow the user to hover on a calibration curve, which displays a tooltip with salient information about the selected set of predictions, like expected calibration error. Furthermore, the user may click to ``select'' the calibration curve, which identifies which model to consider for the brushing operation which populates the Instance and Performance views. With a curve selected, users may also brush to select a prediction region of interest. Using the x-axis values of the brushed region, we then populate the Instance and Performance views with predictions that are within the brushed region. 

Each interviewed expert mentioned that they often coupled a plot of the prediction distribution with the reliability diagram, as it helped to identify areas with high uncertainty due to low sample size. At the bottom of the selected curve, Calibrate provides a histogram of predictions for the observations used to create the selected curve. Understanding the distribution of predictions is important since, especially for difficult multiclass prediction problems, models may not even produce predictions with high confidence. Furthermore, with a density plot, users will be able to gauge the uncertainty of a prediction region, as regions with small amounts of samples will generally have more uncertain estimates.

\subsection{Feature View}
Subgroup analysis is an important yet arduous task for most users. To analyze subgroup calibration, users first define subgroups of interest, often on a single feature, but potentially on groups of features. Typically, these actions are performed manually in Jupyter Notebooks cells. Thus, with our Feature View (Figure~\ref{fig:teaser}-E), we seek to make the subgroup creation process interactive to enable quick and iterative analysis. P1 mentioned that they typically view subgroups in terms of ``distributions''. Thus, we provide a view for users to create subgroups through brushing feature distributions (\ref{req:subgroup}). Being able to visualize the distributions of features may also help practitioners understand the limitations of their own data. For example, P1 mentioned that through exploratory data analysis on their features, they sometimes find important subgroups for which they may lack data, and thus they investigate the calibration characteristics a subgroup defined in this low density region.

In the Feature View, a user must first select what features to visualize. Users may brush each histogram by dragging across a region of the histogram. Once a user has brushed a region, the user then can create a new reliability diagram. The samples that fit the brushed selection will populate a reliability diagram in the calibration view. One may one-hot encode categorical variables, therefore making subgroup creation a matter of brushing either the $0$ or $1$ part of the feature's domain. Within the Feature View, a user can scroll to see the full list of features.

% We noted in initial designs that simply visualizing the distribution of each feature and allowing the user to scroll through each is cumbersome, especially for problems with large feature sets. Therefore, we allow a user to populate the Feature View with histograms of features by selecting features through a drop down menu.

\subsection{Instance View}
Exploring errors on the instance-level is an important aspect of model performance analysis. However, due to the binning required, along with the static nature of reliability diagrams, instance-level inspection is usually not performed in calibration analysis. Furthermore, it is unclear how to assess calibration error on the instance level -- two instance may differ in true and predicted labels, but assessing the difference between predicted probability and outcome is akin to a proper scoring rule, which is why binning is employed to measure calibration. Thus, it is difficult to connect performance (confidence regions) to particular instances. Nevertheless, inspecting individual instances manually can still yield interesting outcomes, and may guide model improvement. To satisfy \ref{req:instance}, we implement the Instance View (Figure~\ref{fig:teaser}-D). As users brush on the calibration view, the instance view updates to reflect the instances in the selected prediction rage. Each row contains information on the features of each instance and the mean of each feature is shown at the bottom of the instance view.

\subsection{Performance View}
Ultimately, confusion matrix-derived metrics are still important to visualize for model performance analysis, and was a standard procedure among the experts we interviewed. Thus, it is important to present these classic model performance measures alongside calibration performance measures. Therefore, we implement the Performance View (Figure~\ref{fig:teaser}-F), which visualizes a confusion matrix, which forms the basis of many count-based performance metrics. The performance view is updated as users brush on the calibration view to select confidence ranges of interest or when a user selects another calibration curve.

\subsection{Implementation} \label{sec:implementation}
Given many data scientists' background in Jupyter, we provide a solution compatible with Jupyter notebooks (\ref{req:jupyter}). The main justification in providing a Jupyter-first tool was that, from our expert interviews, practitioners much preferred to keep their analysis within Jupyter notebooks. In fact, none of our interviewed experts relied on outside programs to analyze their models. The two main avenues of model analysis were either (1) combine model training and performance analysis into a single notebook or (2) model training, particularly for large models, was offloading to another resource, and a notebook read in the predictions for model analysis.

Calibrate is available as a Python library and is designed with Jupyter Notebooks use in mind. The front-end is implemented via JavaScript with React and D3~\cite{DBLP:journals/tvcg/BostockOH11}. The back-end, which creates reliability diagrams, is implemented in Python using Numpy~\cite{harris2020array}, Pandas~\cite{mckinney-proc-scipy-2010} and Interpret~\cite{DBLP:journals/corr/abs-1909-09223}. We use Interpret for its EBM implementation with its default parameters. To use Calibrate, a user first creates the \texttt{Calibrate(data)} class, where \texttt{data} is a dataframe with rows as observations and columns as features. To add models to the Calibrate, users use the \texttt{.add\_model(preds, labels, model\_name)} method, where \texttt{preds} is an $N \times K$ matrix of predicted probabilities, \texttt{labels} is a $N \times K$ one-hot encoded matrix representing the labels of the observations in \texttt{data}, and \texttt{model\_name} is a string indicating the model name. $K$ is equal to the total number of classes. Finally, to visualize the Calibrate widget, a user may call the \texttt{.visualize()} method. 

%  and PyGAM~\cite{serven2018pygam}
%  and PyGAM for its spline-based GAM implementation
% make images for showing how brushing affects instance view and performance view
% make images showing feature view and making a subgroup with conventional knowledge
\section{Evaluation} \label{sec:evaluation}
 \subsection{Case Study 1: Identifying Miscalibration in Subgroups} \label{sec:compas-use-case}
Analyzing model calibration among subgroups is an important task. Oftentimes, discovering that a model is miscalibrated for specific subgroups leads to important business or technical decisions. For example, if a sports bettor discovers their game outcome model is miscalibrated for games that occur at night, they may decide not to bet on such games. Like our experts indicated, if a machine learning practitioner uncovers miscalibration among subgroups in their model, they may decide to engineer more features or to change their model architecture. In this use case, we show how Calibrate can be used to analyze model miscalibration among subgroups in a real world data set.

The COMPAS (Correctional Offender Management Profiling for Alternative Sanctions) data was released by ProPublica in 2016 and is based on Broward County data from January 2013 to December 2014~\cite{angwin2016machine}. The data consists of information about defendants, such as their sex, race, or previous arrests, which are used to create models to predict a defendant's risk of recidivism. These models have increasingly come under scrutiny, due to their ability to cause significant harm to individuals. While the issue of whether or not these models should be used in practice is best left to other domains, understanding recidivism model calibration is nonetheless important and provides a unique example of how miscalibration may harm individuals.

\begin{figure}
    \centering
    \includegraphics[scale=0.63]{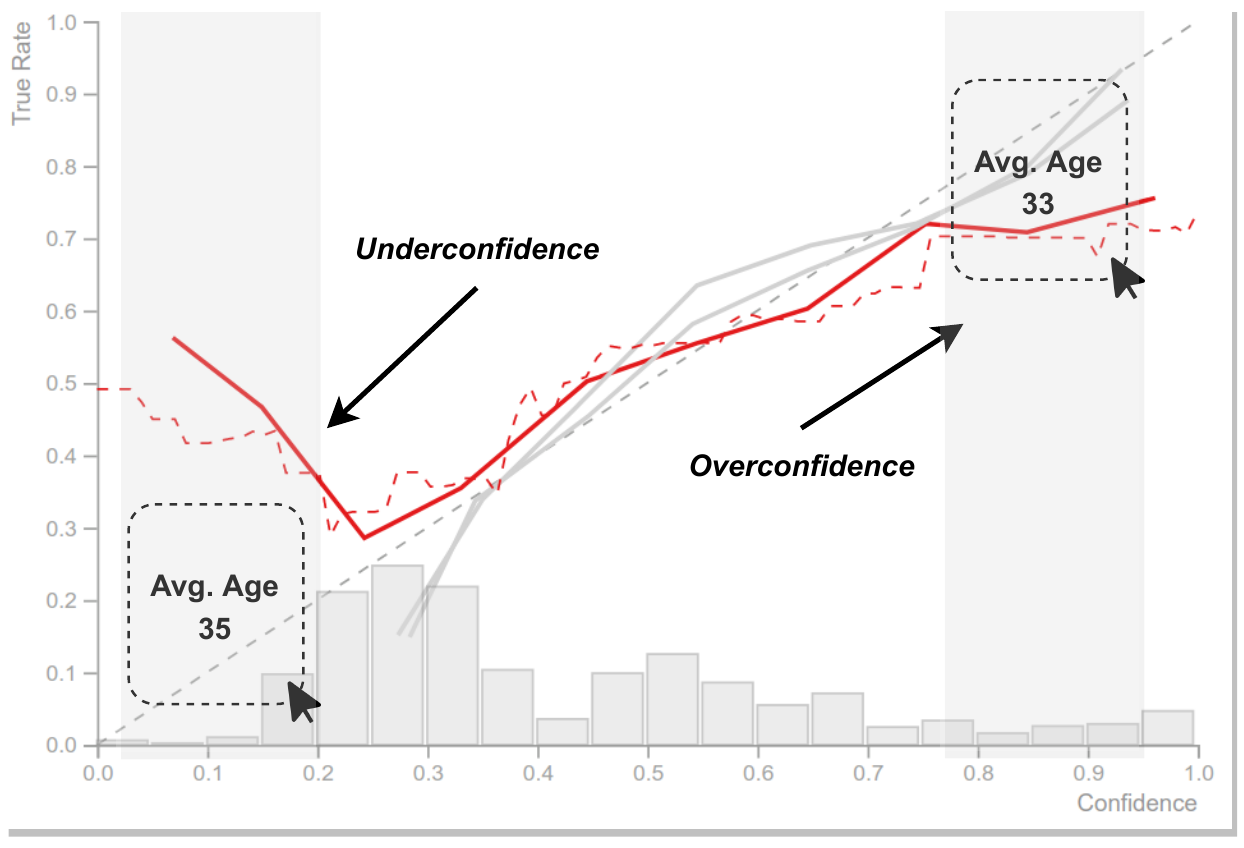} % compas_without_age.pdf, uc_1_original.pdf
    \caption{ \edit{Reliability diagrams for each model, which consider a defendant's sex, charge type, and prior offense counts, on the full COMPAS test set on a model considering. We can see significant regions of under and overconfidence in the random forest model (red, selected). The random forest's learned reliability diagram is shown as the red dotted line, and its histogram of the predicted probabilities is shown along the x-axis. Brushing indicates regions of predictions with above and below average age through the instance view.} }
    \label{fig:compas}
\end{figure}

We split our data with 50\% of observations going to training and 50\% going to testing. To start, we consider a defendant's sex, charge degree (misdemeanor or felony), juvenile misdemeanor and felony counts, as well as total prior charge count. We train a random forest classifier, which was noted by our interviewed practitioners as a common model choice they use in practice. Additionally, we consider a multilayer perceptron (MLP) and a logistic regression as other candidate models. We use the default parameters and implementation in sklearn for each model. Using Calibrate, we show the model's associated reliability diagram, along with its learned reliability diagram, in Figure~\ref{fig:compas}.

% image of full (rf), lr and mlp

\begin{figure*}
    \centering
    \includegraphics[scale=0.63]{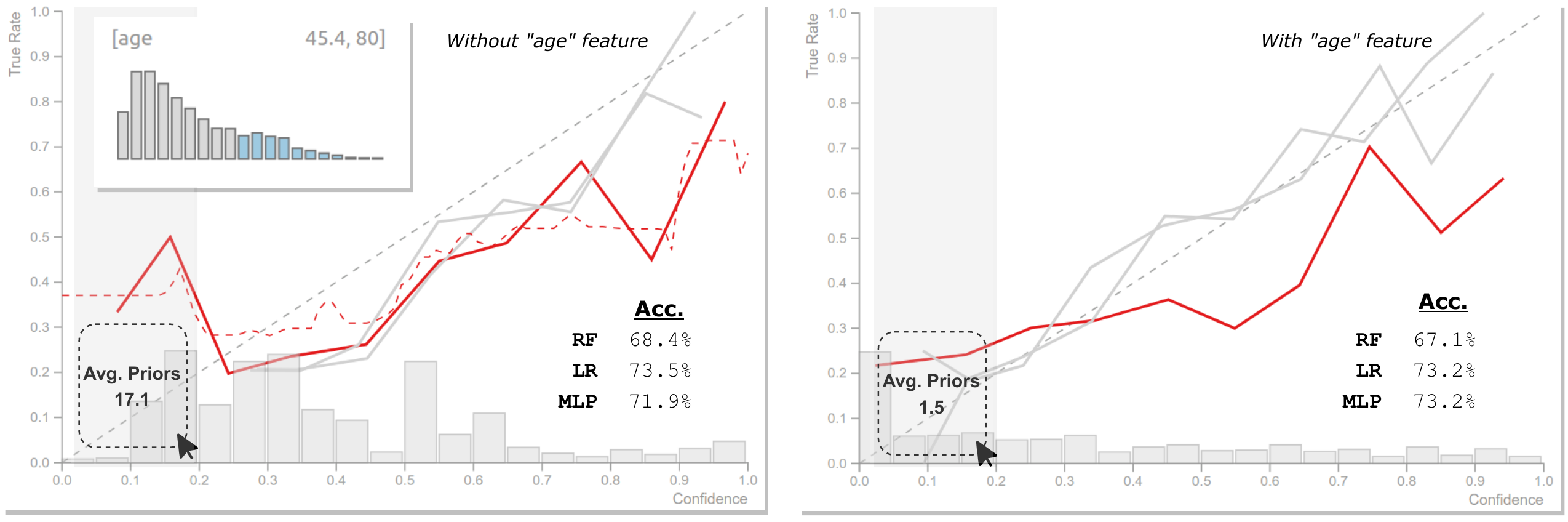} % compas_old_analysis.pdf, uc_1_subgroup.pdf
    \caption{ \edit{Reliability diagram for ``old age'' subgroup -- defendants in COMPAS data who are over the age of 45. Left panel shows models \textit{without} age as a feature, and the right panel shows models \textit{with} age as a feature. We see systemic overconfidence in the predictions for all models that do not consider age as a feature. Histogram of random forest predictions are shown in each panel.} }
    \label{fig:compas-old-age}
\end{figure*}
% This systemic overconfidence was fixed by adding defendant age as a feature, as well as changing from a random forest to a logistic regression or MLP.

\edit{Figure~\ref{fig:compas} shows that the models are well-calibrated, with the exception of our random forest model in the very low and very high predictions, where we see under and overconfidence, respectively. Overconfidence is indicated by values of the reliability diagram below the $y=x$ perfect calibration line. These are samples which are generally predicted as having a higher probability than the true occurrence rate. Conversely, underconfidence implies that the model predicts a lower probability than the true occurrence rate. For our domain, we may be especially concerned with model overconfidence, as such a characteristic results in higher predicted probabilities than the true rate, meaning a defendant will be rated as higher risk than they truly are, on average.}

\edit{While investigating global calibration is typically the beginning of any calibration analysis, it is also important to check a model's calibration characteristics across subgroups of interest. In practice, one would perform subgroup analysis by using domain knowledge to manually define subgroups, through packages like Pandas~\cite{mckinney-proc-scipy-2010}. Analyzing regions of the global reliability diagram can also be a useful direction to identify subgroups. Using the brushing interaction, we analyze both the underconfident and overconfident regions in Figure~\ref{fig:compas}. Interestingly, through the instance view, we observe many young defendants in the overconfident region, which has an average age of 33. In this region, the average predicted probability is about 0.9, yet the average true rate is only about 0.75. Conversely, we see that the underconfident region is slightly older, with an average age of 35. At the same time, at the extremely low end of the underconfident region (i.e., $\textrm{prediction} \leq 0.1$), the average age is just 31.}

\edit{Thus, it is reasonable to posit that the miscalibration, particularly in the low and high predictions, may be correlated to some degree with defendant age. Therefore, we define a subset of data encompassing older defendants, which we define as those of $45+$ years of age. Using Calibrate's feature view, we create the associated reliability diagram for the ``old'' subset in Figure~\ref{fig:compas-old-age}. Interestingly, we see that this group faces systematic overconfidence in its predictions, regardless of the selected model choice. Many of our interviewed experts noted that when faced with model miscalibration, it was common for them to hypothesize what features they may be missing. In this instance, it is reasonable to assume that, since we do not include age as a feature, that we may be observe miscalibration associated with age. Thus, we may also include a defendant's age in our model and observe the resulting reliability diagram in Figure~\ref{fig:compas-old-age}. We see that the models become more calibrated on the old age subgroup, yet suffer no large deviation in accuracy. Interestingly, we also see that the characteristics of the lower 20\% of predictions changes. For example, without ``age'' as a feature, these predictions are about half male and half female with an average of 17 prior arrests. However, when considering age, this subset of predictions drops to an average of about 1.5 prior arrests. Thus, the instance view can be useful for learning about the characteristics of prediction regions.}

\edit{P1 noted that they sometimes noticed miscalibration with certain model choices. For example, P1 stated ``we often use logistic regression as we noticed its outputs were well-calibrated for many prediction tasks''. Aside from feature engineering, another avenue one may consider exploring is model architecture. While this may seemingly address global model calibration issues, as demonstrated in Figure~\ref{fig:compas}, subset analysis is still important. For example, consider the right panel of Figure~\ref{fig:compas-old-age}, where two models can exhibit similar performance on labels, yet have markedly different calibration characteristics.}

\subsection{Case Study 2: Analyzing Determinants of Calibration} \label{sec:determinants-use-case}
Understanding the factors that may induce miscalibration in a model is important. The experts we interviewed typically hypothesized miscalibration as a consequence of one of two factors: (1) lack of useful features or (2) class imbalance. For example, P4 mentioned that if a model was miscalibrated, they would try to answer questions like ``what features am I missing?'' and ``what is the balance of the classes in the data sets?''. In this use case, we investigate how the aforementioned factors influence model calibration through experiments with synthetic data. We use sklearn's ``make\_classification'' implementation to design a classification task with 10 classes. We generate 20 features for each observation, with 10 of the features as informative and the remaining 10 as noise. We randomly select 50\% of the generated samples for training and the remaining 50\% for testing. We train a multilayer perceptron using the default parameters provided by sklearn.

% sample size: 30000, 10000, 5000
% class imb: 10%, 50%, 90%
% feat: no important features, half important features, all features

\subsubsection*{Feature Set} % One option is to use the feature subset to find the important features
Many experts mentioned that they would turn to feature engineering or to collecting different features when faced with model miscalibration. For example, P1 pointed to a real-life use case where they identified a subgroup that was poorly calibrated, which they fixed by creating bespoke features for the subgroup. Thus, it is important to understand the effect of the feature set on model calibration. We also visually demonstrate how, somewhat paradoxically, a model with only uninformative features is often calibrated. We consider a classification task on 10,000 total observations where one class comprises 50\% of all observations, and the remaining 50\% of observations are spread equally among the other 9 classes.

\begin{figure}
    \centering
    \includegraphics[width=\linewidth]{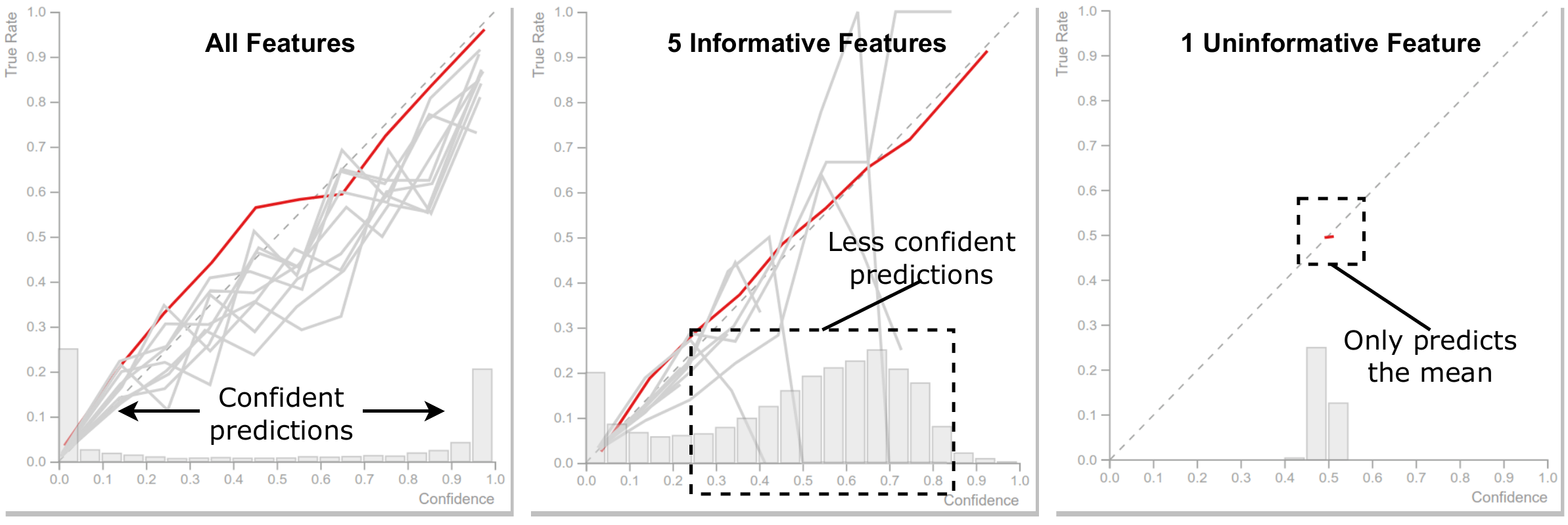} % uc_2_feat.pdf, feature_use_case.pdf
    \caption{Dominant class reliability diagram by feature set. Models trained on all features or 5 informative features are well-calibrated, although the latter has less confident predictions (further from 0 or 1). Using a single uninformative feature produces a model which purely estimates base class -- this model can deceptively yield zero calibration error (calibration curve shown in dotted box) yet be an uninformed model. Histograms of dominant class predictions shown below.}
    \label{fig:feature_set_use_case}
\end{figure}

\edit{ We consider three feature set scenarios: (1) all features, (2) only 5 informative features, (3) 1 uninformative feature. In Figure~\ref{fig:feature_set_use_case}, we visualize the calibration characteristics of the most prevalent class in red, along with the minority classes each as a gray line. We see that the model is well-calibrated when all features are used, as well as when only 5 informative features are used. Although our model is well-calibrated in (1) and (2), the model attains a much lower accuracy in scenario (2). Thus, although two models may both be well-calibrated, shown in the calibration view, such a finding does not guarantee similar accuracy, found in the performance view. Furthermore, using the histogram, we notice that as the predictions become less confident (farther away from 1 or 0) as the number of informative feature decreases. }

For scenario (3), we see a single point around 0.5 confidence and accuracy. Effectively, our model is learning no relationships in this case. Thus, for our dominant class, the model always predicts its base rate -- 50\%. In a sense, this model is very well-calibrated since its expected calibration error will be 0. However, intuitively, this model has low accuracy. Thus, we show a canonical example with zero calibration loss, but high error rate (and thus high log loss). Furthermore, this demonstrates a strong case for the use of reliability diagrams in investigating calibration -- simply calculating expected calibration error would not have led to the same conclusion.

\subsubsection*{Class Imbalance} % Use LRDs and ability to show multiple classes in one-versus-rest fashion
Class imbalance, where a particular class is more prevalent than others (known as the dominant or majority class), is common in real-world data. Consider a simple data set with two classes, A and B, where the ratio of A instances to B is $99:1$. A naive model which always predicts class A as $0.99$ and class B as $0.01$ will achieve 99\% accuracy and \textit{technically} be perfectly calibrated, as we saw above. However, this model would unlikely be used in practice, since it provides no new information to practitioners. Furthermore, imbalanced class scenarios often carry imbalanced costs between errors for different classes, which make identifying minority class instances important. Therefore, it is important to understand class imbalance's relationship with calibration.

\begin{figure}
    \centering
    \includegraphics[width=\linewidth]{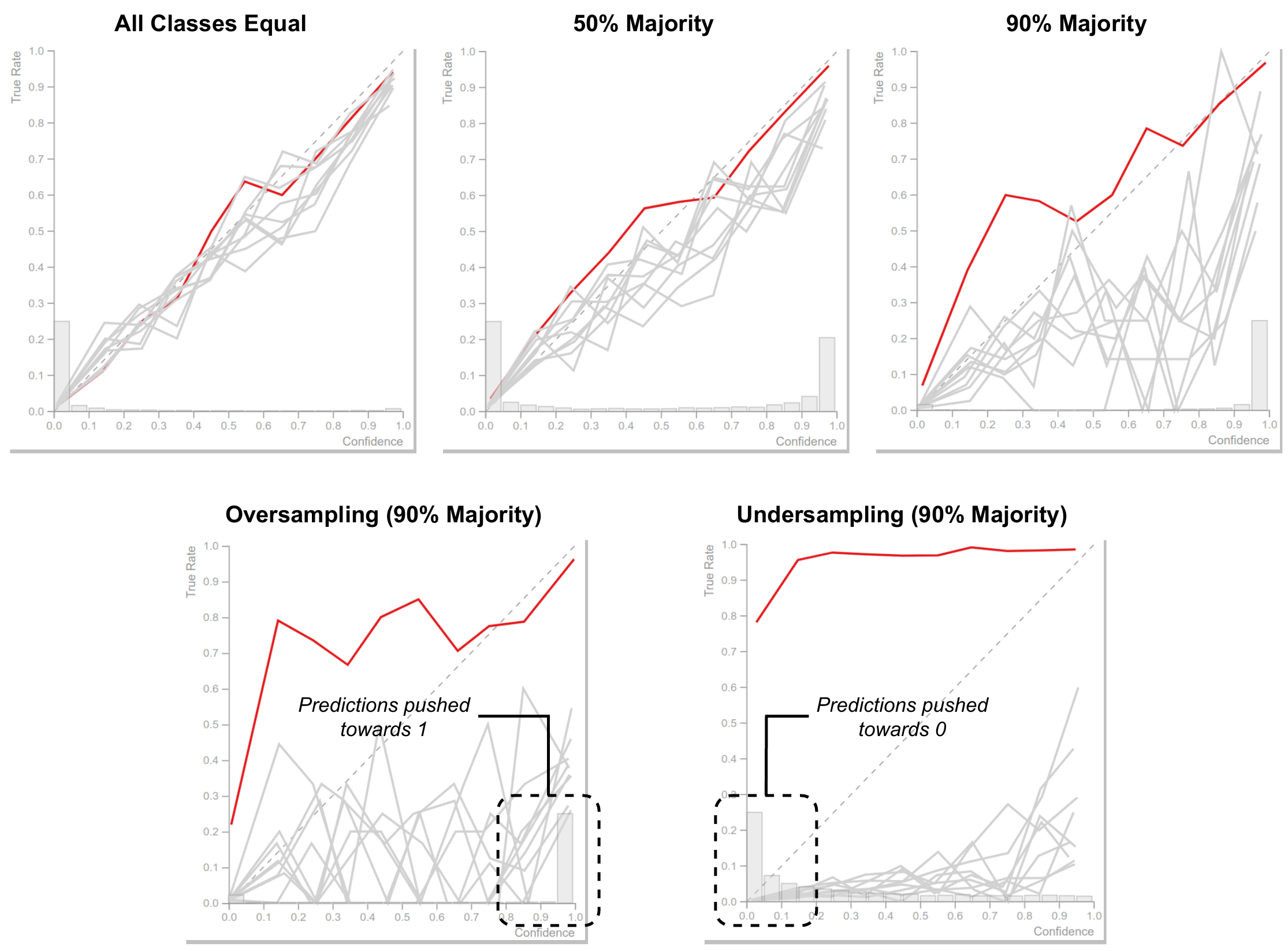} % uc_2_class.pdf
    \caption{ \edit{ Reliability diagram by class imbalance with majority class prediction histogram shown. Each model is well-calibrated for the majority class (red). As class imbalance increases, minority class calibration (gray) worsens, and trends towards overconfidence. Class balancing techniques appear to make model predictions even more miscalibrated.} }
    \label{fig:class_imbalance}
\end{figure}

\edit{We consider three scenarios: (1) equal class balance, (2) dominant class is 50\% of samples, (3) dominant class is 90\% of samples. In total we use 10 classes, and the non-dominant classes consist of equal portions of the remaining samples. In Figure~\ref{fig:class_imbalance}, we show the reliability diagrams for the dominant class (red) and the non-dominant classes (gray), which we created by toggling the selected class in the calibration view. Intuitively, the majority class instances are well-calibrated across the three imbalance levels. However, we see that as class imbalance increases, minority instances become systemically miscalibrated, as indicated by the many gray lines below the 45-degree dotted line. The direction of the miscalibration is also towards overconfidence. Overconfidence implies that the model predicts a class with high probability, yet the true probability, as indicated by the reliability diagram, is low. }

% We consider three scenarios: (1) equal class balance, (2) dominant class is 50\% of samples, (3) dominant class is 90\% of samples. In total we use 10 classes, so the non-dominant classes consist of equal portions of the remaining samples. In Figure~\ref{fig:class_imbalance}, we show the reliability diagrams for the dominant class (blue) and one of the non-dominant classes (red). Intuitively, majority class instances are well-calibrated across all levels of imbalance. However, we see that as class imbalance increases, minority samples become more imbalanced. The direction of imbalance is also towards overconfidence, which we confirm across all minority classes. Overconfidence in this instance implies that the model predicts a class with high probability yet the true probability, as indicated by the reliability diagram, is low.

\edit{ One common approach to address class imbalance is to resample the training data in a way that evens out the class distribution. Undersampling evens the class distribution by sampling only a portion of the majority class so that the sampled number of points equals the total minority sample count. Oversampling resamples the minority classes so that the sampled number of points equals the total majority sample count. We run an experiment using the data from the 90\% majority dataset. We show the results of applying basic over and undersampling in Figure~\ref{fig:class_imbalance}. In our experiment, class-based resampling has significant effects on the resulting diagram, which appeared to make our predictions miscalibrated. Specifically, the majority class predictions became severely underconfident, while minority class predictions became severely overconfident. Furthermore, the prediction histogram shows that oversampling pushed majority class predictions towards 1, while undersampling pushed majority class predictions towards 0.}

\subsection{Expert Feedback}
We conducted ``think aloud'' interviews with the same experts who had previously helped us identify the system requirements. In this session, the experts had access to Calibrate via a Jupyter notebook. The notebook was preloaded with a trained model on the Census Income data set~\cite{Dua:2019}. Before participants interacted with Calibrate, they were given a brief demo of Calibrate's functionality. Participants were free to explore the results of Calibrate in whatever way they wanted.

User feedback was very positive. The experts mentioned that they found the system useful and were interested in using Calibrate for their own work. In particular, the experts appreciated the ability to quickly create subgroups through interaction. P1 mentioned that in the future, it may be useful to have an ordering to the features, such as by their variance. The experts also enjoyed the interaction of selecting a calibration curve, and seeing both the prediction distribution and instances. P2 remarked that the information displayed in Calibrate was clear and interpretable, and that they may consider using the tool for non-technical stakeholders as well. P3 found Calibrate easy to use, especially due to its Jupyter implementation.

P1 mentioned that although the learned reliability diagram was intuitive as a concept, that the shape of the learned diagram was unintuitive. For the given demo, we utilized an explainable boosting machine~\cite{DBLP:journals/corr/abs-1909-09223} for the learned reliability diagram, which resulted in a ``step-like'' curve, according to P1. Because of this, P1 mentioned that they confused the steps in the curve as areas of significant calibration changes. When shown a learned reliability diagram using splines, P1 remarked that the smooth function created by the splines was much more intuitive. 

%likes the filtering BUT keep highlighted filter on feature view
%like the clicking to see the background dist, likes to be able to look thru samples
%make it equal to the number of bins on the reliability curve. see distribution in context of the line
%scroll on feature view.
%confusion matrix not super helpful, bunch of predictions in top left corner. prefer number, hard to tell
%spline vs ebm
%sort features by variance (interesting! like this. easy.)
%super simple with .add_model
%useful to add post-hoc calibration (.calibrate)
\section{Discussion} \label{sec:discussion}
One limitation of our work is that we do not directly tackle the issue of multiclass calibration visualization, and we opt for a one-versus-rest approach instead. In part, this design choice is due to the ubiquity of traditional reliability diagrams, which appear to be easily understood by both technical and non-technical audiences. Furthermore, it was unclear from our domain requirements interviews that multiclass calibration was commonly analyzed, and if so, a one-versus-rest approach was often deployed. Works like Vaicenavicius~et~al. propose a method to visualize calibration across prediction tasks with three and four classes~\cite{DBLP:conf/aistats/VaicenaviciusWA19}. One solution to multiclass calibration visualization may be to adopt a Squares-like~\cite{DBLP:journals/tvcg/RenALSW17} approach, whereby each class is represented on a single horizontal axis by a vertical reliability diagram. Parallel coordinates could also be used to link bins and instances across classes.
% A downside to their approach is that for tasks with a large class count, classes would needed to be grouped together, and calibration distinctions between classes may be muted.

One limitation of learned reliability diagrams is that the selected model architecture may induce particular reliability diagram shapes. For example, if one uses logistic regression as the basis of their learned reliability diagram, then one would see a sigmoid relationship, which, in some instances, may imply overconfidence in the high probabilities and underconfidence in low probabilities. For the case of EBMs, which we use to produce our learned reliability diagrams, we found that, for low max bin counts, the learned reliability diagram sometimes appeared underconfident at the low end of the predictions. However, as shown in Figure~\ref{fig:lrd}, setting the max bins parameter to a high number has little effect on the shape of the learned reliability diagram. Thus, by using the default parameters of EBMs, one may avoid this problem.

While calibration has traditionally been a topic covered mostly in data mining or machine learning, there are exciting avenues for calibration work in visualization. For example, developing and evaluating new visual encodings for model calibration is particularly interesting. Reliability diagrams rely on many visual design decisions, such as whether to use points or bars, how to visualize uncertainty, or how to set the number of bins. Furthermore, as we saw in our expert study, some users may find certain learned diagram shapes confusing. Many of the questions regarding the aforementioned design decisions may ultimately best be examined through user studies. \edit{Indeed, the evaluation of Calibrate is another limitation of our work. While we primarily demonstrate Calibrate's effectiveness through case studies and expert interviews, a quantitative user study, like that performed by Sahann~et~al.~\cite{DBLP:conf/visualization/SahannMS21}, would be helpful to further demonstrate the effectiveness of Calibrate's features in aiding machine learning model calibration analysis. Furthermore, we note that our presented use cases represent starting points to generate hypotheses, and should be followed by deeper dives.}

%likes the filtering BUT keep highlighted filter on feature view
%like the clicking to see the background dist, likes to be able to look thru samples
%make it equal to the number of bins on the reliability curve. see distribution in context of the line
%scroll on feature view.
%confusion matrix not super helpful, bunch of predictions in top left corner. prefer number, hard to tell
%spline vs ebm
%sort features by variance (interesting! like this. easy.)
%super simple with .add_model
%useful to add post-hoc calibration (.calibrate)
\section{Conclusion} \label{sec:conclusion}
Machine learning model performance analysis is an important task that influences real-life model deployment. Much of current model performance analysis is centered around analyzing a model's predicted labels. However, many applications rely on predicting correct \textit{probabilities} rather than labels. When a model's predicted probabilities align with reality, the model is said to be calibrated. Typically, calibration is explored through static visualizations called reliability diagrams. Since models can achieve high accuracy yet be poorly calibrated, it is important to analyze model calibration. In this work, we present Calibrate, an interactive tool to analyze model calibration. Additionally, we introduce Learned Reliability Diagrams, a simple approach to address the issues with standard binning procedures for traditional reliability diagrams. Calibrate allows users to perform a wide array of analyses, such as subgroup and instance-level analysis. We implement Calibrate for easy use within Jupyter Notebooks, so that calibration analyses can be easily created, reproduced and disseminated. To evaluate Calibrate, we present use cases which perform subgroup analysis and explore model calibration factors, along with expert interviews.

\acknowledgments{
This collaboration has been supported by a grant from Capital One. Silva’s research
has also been supported by NASA; NSF awards CNS-1229185, CCF1533564, CNS-1544753, CNS-1730396, CNS-1828576, CNS-1626098; and DARPA PTG and D3M. Nonato's research has been supported by São Paulo Research Foundation (FAPESP)-Brazil (grant 2013/ 07375-0) and CNPq-Brazil (grant 307184/2021-8). Any opinions, findings, and conclusions or recommendations expressed in this material are those of the authors and do not necessarily reflect the views of DARPA, NSF, NASA, FAPESP, CNPq, or Capital One.
}

\bibliographystyle{abbrv}

\bibliography{00_bib}
\end{document}